\documentclass[%
aps,prf,
 preprint,%
longbibliography,
]{revtex4-1}

\usepackage{amsmath}
\usepackage{amssymb}
\usepackage{mathrsfs}
\usepackage[scr=rsfs,cal=boondox]{mathalfa}

\usepackage{dcolumn}
\usepackage{bm}
\usepackage[pdftex]{graphicx}
\usepackage{epsfig}
\usepackage{epstopdf}
\usepackage{lineno}

\usepackage{color}

\begin{document}
\title{Scaling patch analysis of turbulent kinetic energy budget equation in wall-bounded flows}
\author{Tie Wei}
 \altaffiliation[]{Department of Mechanical Engineering, New Mexico Institute of Mining and Technology.}
 \email{tie.wei@nmt.edu}
\author{Zhaorui Li}
 \altaffiliation[]{Department of Engineering, Texas A\&M University-Corpus Christi, 6300 Ocean Drive, Corpus Christi, TX 78412, USA}
 \email{Zhaorui.Li@tamucc.edu} 
\author{Sergio Pirozzoli}  
 \altaffiliation[]{Dipartimento di Ingegneria Meccanica e Aerospaziale, Universit\'a di Roma ‘La Sapienza’,
 	Via Eudossiana 18, 00184 Roma, Italy}
 	\email{sergio.pirozzoli@uniroma1.it }
 
\date{\today}

\begin{abstract}

The scaling patch approach is applied to analyze the turbulent kinetic energy (TKE) budget equation in wall-bounded turbulent flows. The balance of the TKE equation is divided into several distinct regions, or \textit{scaling patches}, each characterized by a dominant balance among the governing terms and its own appropriate scaling parameters. In the near-wall viscous sublayer, the TKE balance is primarily between viscous diffusion and dissipation, and the characteristic scales are set by the kinematic viscosity $\nu$ and the wall dissipation rate $\epsilon_\mathrm{wall}$. The thickness of this sublayer is on the order of the Kolmogorov length scale.  Moving away from the wall, the peak TKE production provides a natural reference scale for the inner layer, yielding the traditional inner scaling of $u_\tau^4 / \nu$, where $u_\tau$ is the friction velocity. Grouping the viscous diffusion and dissipation terms in the inner layer enhances the collapse across different Reynolds numbers. In the outer region, Prandtl’s mixing-length model is used to derive a characteristic scale for TKE production as $u_\tau^3 / \delta_c$, where $\delta_c$ is the channel half-height. A new meso-scaling is further introduced to describe the intermediate region, ensuring a smooth transition between the inner and outer layers. The scaling patch framework offers a unified interpretation of the structure and scaling behavior of the TKE budget across all regions of wall-bounded turbulence.

\end{abstract}

\maketitle

\section{Introduction}

The turbulent kinetic energy (TKE) budget equation provides a fundamental framework for understanding the generation, transport, and dissipation of turbulence in wall-bounded and free-shear flows. It expresses the balance among production, turbulent and viscous transport, pressure diffusion, and dissipation—each representing a distinct physical process in the turbulence energy cycle. This balance links mean flow gradients to velocity fluctuations, connecting large-scale energy input to small-scale viscous dissipation. Analyzing the TKE budget reveals how turbulence adapts to varying flow conditions such as pressure gradients, wall roughness, or compressibility, and identifies regions where different mechanisms dominate. As such, the TKE budget serves as a key diagnostic for developing scaling laws, advancing turbulence modeling, and deepening physical understanding of turbulent flow structure and dynamics.

The physical interpretation of the TKE budget and its constituent terms has been extensively discussed in classical turbulence literature. The comprehensive work of Tennekes and Lumley \cite{tennekes1972first} remains a cornerstone reference, providing a lucid exposition of the roles of production, transport, pressure diffusion, and dissipation in maintaining the turbulent energy balance. Townsend \cite{townsend1956} further advanced this understanding by emphasizing the influence of large-scale coherent structures on energy transfer processes, particularly in wall-bounded flows. His analysis revealed how organized motions in the outer region modulate the production and redistribution of turbulence energy, establishing a physical link between structural dynamics and statistical energy balances.

Before the advent of high-fidelity numerical simulations, accurate quantification of TKE budget terms was severely constrained by experimental limitations, especially in measuring small-scale dissipation and pressure-related terms. A major breakthrough came with the direct numerical simulations (DNS) of turbulent channel flow by Moser, Mansour and coworkers \cite{moser1999direct, mansour1988reynolds} which provided the first comprehensive, term-by-term evaluation of the TKE budget. Their dataset enabled systematic analysis of the wall-normal distribution of production, dissipation, turbulent transport, and pressure transport, and has since become a benchmark for both turbulence model validation and scaling analyses of individual budget terms.

Subsequent DNS studies at higher Reynolds numbers extended this foundational work and clarified scaling behaviors across different wall-normal regions. Hoyas and Jiménez \cite{hoyas2008reynolds} examined the Reynolds-stress and TKE budgets in turbulent channel flows and demonstrated that, for $y^+ \gtrsim 150$, most budget terms exhibit robust outer scaling with $u_\tau^3/\delta_c$, where $u_\tau$ is the friction velocity and $\delta_c$ is the channel half-height. Here $y^+ = y u_\tau/\nu$ is the inner scaled distance from the wall. However, they also observed that classical inner scaling with $u_\tau^4/\nu$ performs poorly in the buffer layer, particularly for the dissipation and pressure-related terms, attributing this deviation to the influence of large-scale wall-parallel motions.

More recent large-scale DNS efforts by Bernardini et al. \cite{bernardini2014velocity} and Lee and Moser \cite{lee2015direct} further extended the accessible Reynolds-number range, providing carefully processed TKE budgets that allowed for systematic assessment of scaling hypotheses.

Together, these studies have established a robust foundation for interpreting and modeling the turbulent kinetic energy budget. Nevertheless, despite decades of progress, a unified framework that consistently delineates the appropriate scaling for each region of the flow remains an open challenge. The present study addresses this gap by applying the scaling patch approach \citep{fife2005multiscaling, wei2005properties, fife2005stress, fife2006_review, fife2009time, wei2020scaling} to the TKE budget equation, enabling a systematic identification of subdomains in which distinct dynamical balances and scalings arise naturally from the governing equations.

To demonstrate the framework, we first analyse the TKE budget in a fully developed turbulent channel flow, which serves as a canonical configuration with well-defined boundary conditions and minimal geometric complexity. The conventional scaling of the budget terms is introduced in Section~\ref{section:traditional}, followed by a systematic scaling analysis across the flow regions in Section~\ref{section:channel}. The same methodology is subsequently applied to the turbulent boundary layer over a flat plate in Section~\ref{section:ZPGTBL}.

\section{Traditional presentation of TKE budget equation in  turbulent channel flow}\label{section:traditional}

For a fully developed incompressible turbulent channel flow, the  TKE budget equation is expressed as \cite{tennekes1972first, pope2001turbulent}
\begin{equation}\label{eq:TKE_budget}
0 =  -\overline{u'v'} \, \frac{\partial \overline{u}}{\partial y} + \nu \frac{\partial^2 k}{\partial y^2} - \nu\, \overline{ \frac{\partial u'_i}{\partial x_j} \frac{\partial u'_i}{\partial x_j}}    - \frac{\partial}{\partial y} \Big( \frac{1}{2} \overline{u'_j u'_j v'}  \Big) - \frac{1}{\rho}\,\frac{\partial}{\partial y} \Big(  \overline{p' v'} \Big),
\end{equation}
where $\rho$ and $\nu$ denote the fluid density and kinematic viscosity, respectively, and $p$ is the pressure. The streamwise and wall-normal directions are represented by $x$ and $y$, respectively. Overbars denote Reynolds averaging, and primed quantities represent fluctuations about the mean. For instance, $\overline{u}$ is the mean streamwise velocity, $u'$ and $v'$ are the streamwise and wall-normal velocity fluctuations, respectively. The turbulent kinetic energy is defined as $k = \tfrac{1}{2}(\overline{u'u'} + \overline{v'v'} + \overline{w'w'})$.

In Eq.~\eqref{eq:TKE_budget}, the first term on the right-hand side represents the production of TKE due to mean shear. The second and third terms account for molecular diffusion and viscous dissipation (also denoted as $\epsilon$ for brevity), respectively. The fourth term corresponds to turbulent transport, while the final term represents transport associated with pressure fluctuations.

In the traditional presentation of TKE budget equation (\ref{eq:TKE_budget}), friction velocity $u_\tau$ is taken as the characteristic velocity, and the friction length $\nu/u_\tau$ is used as the characteristic length scale. The traditional inner scaled variables are denoted as 
\begin{equation}
y^+ = \frac{y}{\nu/u_\tau}; \quad \overline{u}^+ = \frac{\overline{u}}{u_\tau}; \quad -\overline{u'v'}^+ = \frac{-\overline{ u' v'} }{u^2_\tau}; \quad k^+ = \frac{k}{u^2_\tau}; \quad \overline{p'v'}^+ = \frac{\overline{p'v'}}{\rho \, u^3_\tau}.
\end{equation}
Hence, the traditionally  inner-scaled TKE budget equation can be presented as
\begin{equation}\label{eq:traditional_inner_scaling}
0 =  -\overline{u'v'}^+ \, \frac{\partial  \overline{u}^+}{\partial  y^+} +   \frac{\partial^2 k^+}{\partial  (y^+)^2} - \overline{ \frac{\partial u'^+_i}{\partial x^+_j} \frac{\partial u'^+_i}{\partial x^+_j}}    - \frac{\partial }{\partial y^+} \Big( \frac{1}{2} \overline{u'_j u'_j v'}^+  \Big) - \frac{\partial }{\partial y^+} \Big(  \overline{p' v'}^+ \Big).
\end{equation}

\begin{figure}[h]
	\centering
	\vspace{3pt}
	\centerline{\hbox{ \hspace{-.0in} 
			\epsfxsize=6in
			\epsffile{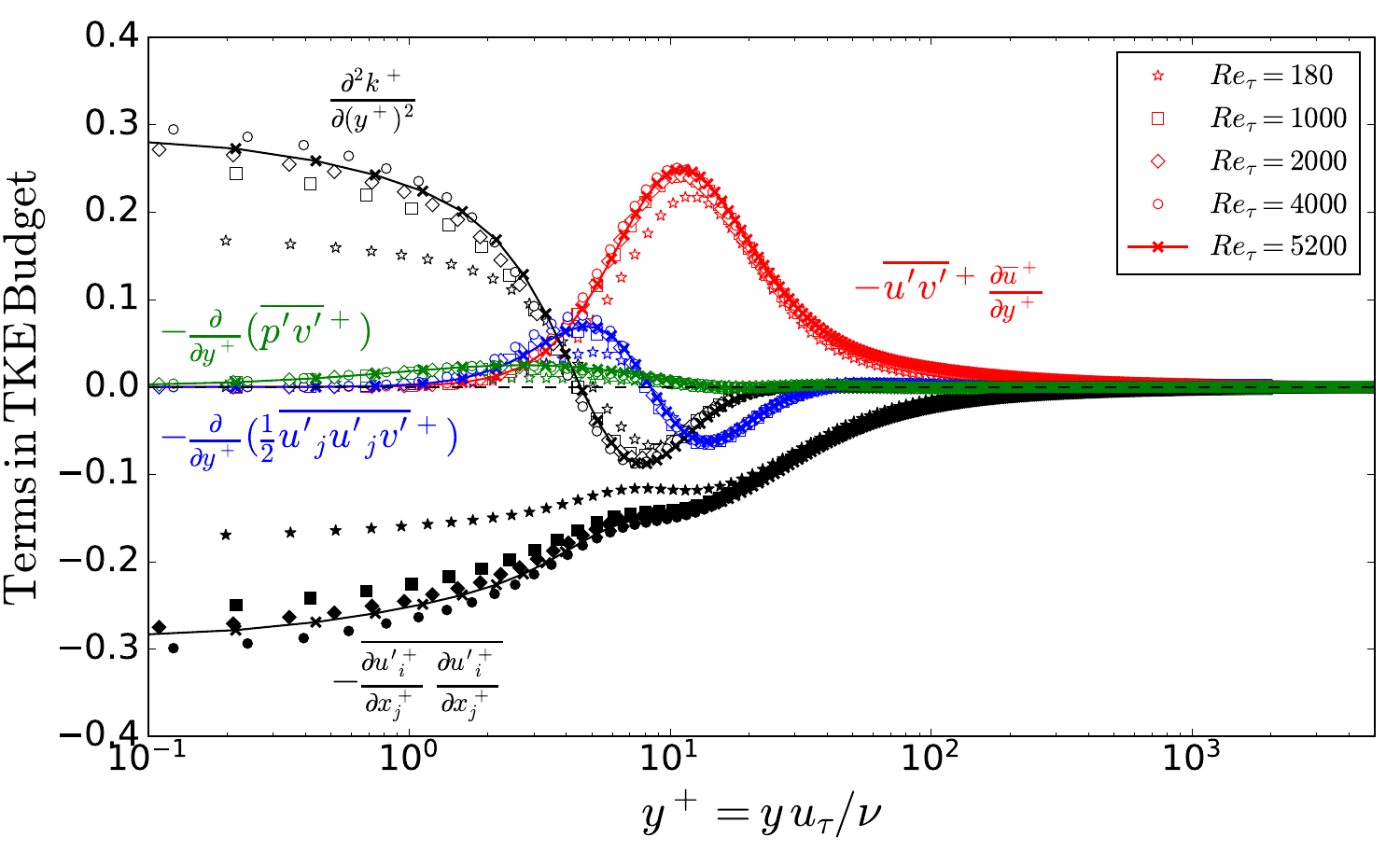}
		}
	}
	\vspace{-10pt}

	\caption{\it Traditional presentation of the terms in the TKE budget equation for turbulent channel flow. Data at $Re_\tau = u_\tau \delta_c / \nu = 180$, $1000$, $2000$, and $4000$ are taken from the DNS of Bernardini et al.~\cite{bernardini2014velocity}, while data at $Re_\tau = 5200$ are from the DNS of Lee and Moser~\cite{lee2015direct}. }
	\label{fig:traditional-presentation-TKE-budget}
\end{figure}

Figure~\ref{fig:traditional-presentation-TKE-budget} presents the traditionally inner-scaled TKE budget equation~(\ref{eq:traditional_inner_scaling}) for five Reynolds numbers ranging from $Re_\tau = 180$ to $5200$. It is evident that this conventional scaling fails to accurately capture the near-wall behavior, where the inner-scaled viscous diffusion and dissipation exhibit a clear Reynolds-number dependence. Moreover, the same scaling proves inadequate in the outer region, where all normalized quantities are much smaller than unity. To overcome these limitations, we employ the scaling-patch framework to identify the appropriate scaling of the TKE budget equation within the distinct layers—or patches—of the turbulent flow.

\section{Application of the scaling patch framework to the TKE budget in turbulent channel flow}\label{section:channel}

In many physical systems, processes occur across multiple spatial and temporal scales, often governed by parameters that are either very small or very large. To describe such multi-scale systems effectively, it is advantageous to introduce rescaled variables—transformations of the original variables that depend explicitly on these parameters. A rescaling is considered natural for a given subdomain if, when expressed in terms of the new variables, the dependent quantities remain of order unity ($O(1)$) and vary at moderate rates; that is, their scaled derivatives are bounded and neither excessively large nor negligibly small. This principle, central to the concept of scaling patches \cite{fife2005multiscaling, fife2006_review}, provides a systematic means of partitioning complex multi-scale phenomena into regions where distinct dominant balances prevail, thereby enabling clearer interpretation and more effective analysis of the underlying physical processes.

To illustrate the scaling-patch approach procedure, we begin with a generic balance equation of the form
\begin{equation}
	0 = A + B + C + D + E,
\end{equation}
where $A$, $B$, $C$, $D$, and $E$ represent different contributing terms. In the context of fluid-thermal systems, each term may involve velocity components, pressure, temperature, and spatial derivatives. In conventional scaling analysis, one typically identifies characteristic scales for each component within the terms $A$–$E$. For instance, consider the production term in the TKE budget equation. The conventional inner scaling sets the reference for the Reynolds shear stress $-\overline{u'v'}$ as $u_\tau^2$, for the mean velocity $\overline{u}$ as $u_\tau$, and for the length scale as the viscous length $\nu/u_\tau$. Consequently, the reference scale for the entire TKE production term becomes $u_\tau^4/\nu$. However, when a term involves multiple interacting quantities—such as in the Poisson equation (see Ref.~\cite{wei2025scaling})—it can be difficult to identify suitable scales for each component.

In the scaling-patch approach used in Wei and Pirozzoli \cite{wei2025scaling}, instead of assigning separate scales to each component, they introduce a lumped reference scale for each term within a given patch:
\begin{equation}
	A^* = \frac{A}{A_\mathrm{ref}}; \quad B^* = \frac{B}{B_\mathrm{ref}}; \quad C^* = \frac{C}{C_\mathrm{ref}}; \quad D^* = \frac{D}{D_\mathrm{ref}}; \quad E^* = \frac{E}{E_\mathrm{ref}}.
\end{equation}
For the appropriate scaling within a patch, all normalized variables $A^*$, $B^*$, $C^*$, $D^*$, and $E^*$ should be of order $O(1)$, or much smaller ($\ll O(1)$) if they make negligible contributions to the local balance. Thus, at least two terms must be of $O(1)$ for a meaningful balance to exist within that patch.

The original equation can then be rewritten in normalized form as
\begin{equation}
	0 = A_\mathrm{ref} A^* \;+\;  B_\mathrm{ref} B^* \;+\;  C_\mathrm{ref} C^* \;+\;  D_\mathrm{ref} D^* \;+\;  E_\mathrm{ref} E^*
\end{equation}
If we assume that term $A$ is important in the balance of the equation within this patch, we may divide through by $A_\mathrm{ref}$, yielding a dimensionless form of the governing balance:
\begin{equation}\label{eq:generic_equation_A}
	0 = A^* \;+\; \frac{B_\mathrm{ref}}{A_\mathrm{ref}} B^* \;+\; \frac{C_\mathrm{ref}}{A_\mathrm{ref}} C^* \;+\; \frac{D_\mathrm{ref}}{A_\mathrm{ref}} D^* \;+\; \frac{E_\mathrm{ref}}{A_\mathrm{ref}} E^*. 
\end{equation}
As an example, if the balance within this patch involves only terms $A$, $B$, and $C$, then the ratios $B_\mathrm{ref}/A_\mathrm{ref}$ and $C_\mathrm{ref}/A_\mathrm{ref}$ must be of order $O(1)$, while $D_\mathrm{ref}/A_\mathrm{ref}$ and $E_\mathrm{ref}/A_\mathrm{ref}$ are $\ll O(1)$ and can thus be neglected. Once a patch and its dominant terms are identified, only one reference scale (e.g., $A_\mathrm{ref}$) is required, which can then be used to normalize all other leading-order terms (e.g., $B$ and $C$). This framework forms the basis of the scaling-patch analysis applied in the present study to the TKE budget equation.

\begin{figure}[h]
	\centering
	\vspace{3pt}
	\centerline{\hbox{ \hspace{-.0in} 
			\epsfxsize=6in
			\epsffile{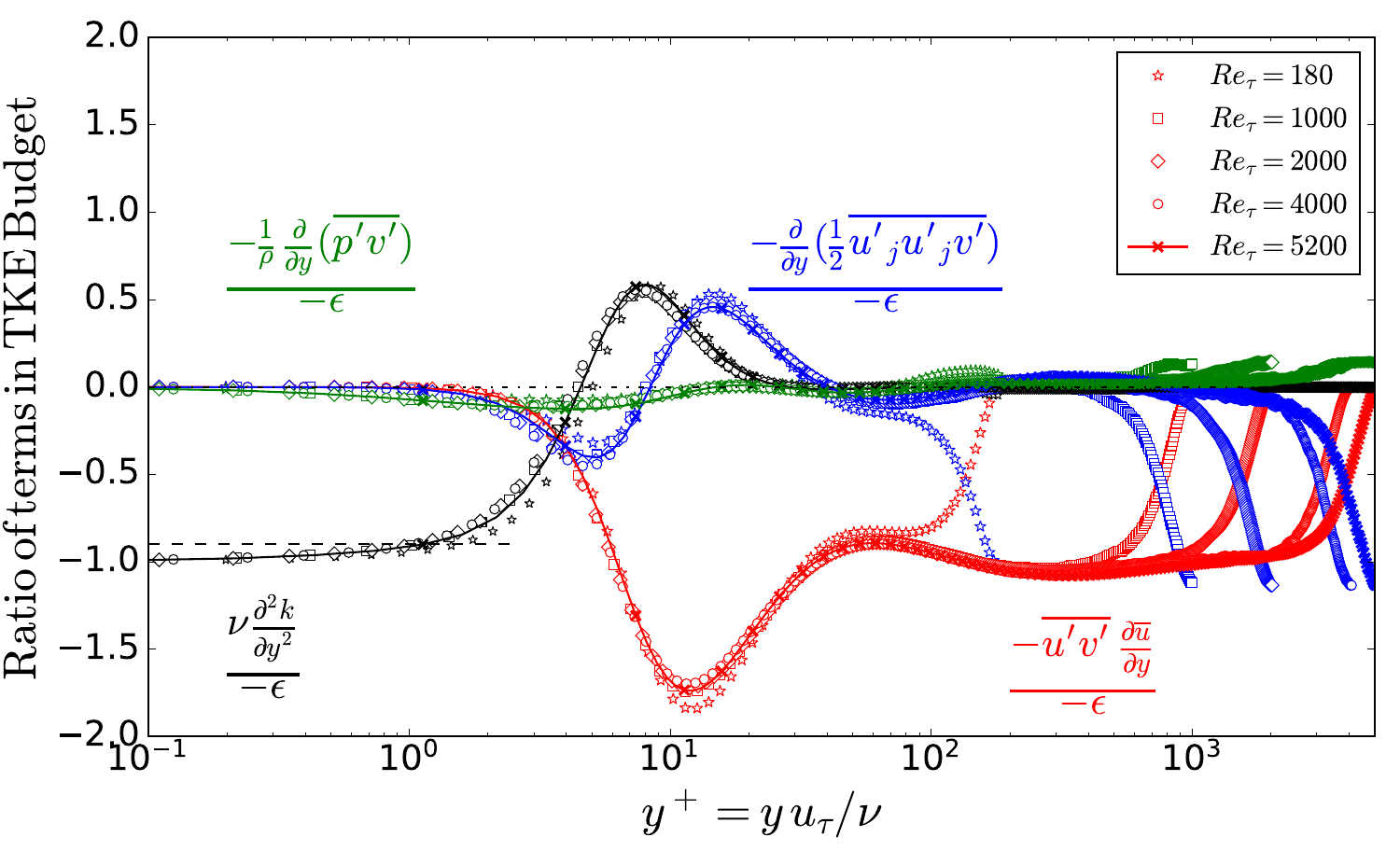}
		}
	}
	\vspace{-10pt}
	\caption{\it Ratio of the terms in the TKE budget equation to dissipation term $\epsilon$. Data sources as in Figure \ref{fig:traditional-presentation-TKE-budget}.  }
	\label{fig:ratio_over_dissipation}
\end{figure}

A practical means of identifying scaling patches is to examine the relative magnitude of the terms in the governing equations. For instance, figure~\ref{fig:ratio_over_dissipation} shows the ratio of each term in the TKE budget equation~(\ref{eq:TKE_budget}) to the viscous dissipation term. The resulting distribution reveals distinct regions, or patches, where different mechanisms dominate the balance. In the near-wall sublayer, viscous diffusion and dissipation are the leading contributors. Moving away from the wall, the dominant balance gradually shifts among dissipation, production, and turbulent transport. Toward the channel centerline, the role of production weakens, while turbulent transport becomes increasingly important in balancing dissipation.

Having identified these regions of distinct dynamic behavior, the next step is to determine the appropriate scale for each patch, which will clarify how the governing balances evolve across the wall-normal direction.

\subsection{Scaling in the viscous sub-layer}\label{section:viscous-scaling}
As shown in figure~\ref{fig:ratio_over_dissipation}, within the near-wall sublayer, the TKE budget is primarily governed by two terms associated with molecular viscosity:
\begin{equation}
	0 \approx \nu \frac{\partial^2 k}{\partial  y^2} - \nu \overline{ \frac{\partial u'_i}{\partial x_j} \frac{\partial u'_i}{\partial x_j}}. 
\end{equation}
Since viscous diffusion and dissipation dominate the energy balance in this region, it is natural to select the wall dissipation rate, $\epsilon_\mathrm{wall}$, as the reference scale $A_\mathrm{ref}$ in the generic scaling-patch formulation~(\ref{eq:generic_equation_A}). This choice ensures that the normalized dissipation varies smoothly between $1$ and $0$ across the layer adjacent to the wall, satisfying the definition of a scaling patch—where the normalized quantities remain of order unity and vary at moderate rates. Accordingly, the dimensionless form of the TKE budget in the patch adjacent to the wall can be expressed as
\begin{equation}\label{eq:scaling-viscous-sublayer}
	0 = \Big(\nu \frac{\partial^2 k}{\partial  y^2} \Big)^* - \epsilon^*,
\end{equation}
where
\begin{equation}
	\Big(\nu \frac{\partial^2 k}{\partial  y^2} \Big)^* = \frac{\nu \frac{\partial^2 k}{\partial  y^2}}{ \epsilon_\mathrm{wall}}; \qquad  \epsilon^* = \frac{\epsilon}{\epsilon_\mathrm{wall}}.
\end{equation}

\begin{figure}[h]
	\centering
	\vspace{3pt}
	\centerline{\hbox{ \hspace{-.0in} 
			\epsfxsize=6in
			\epsffile{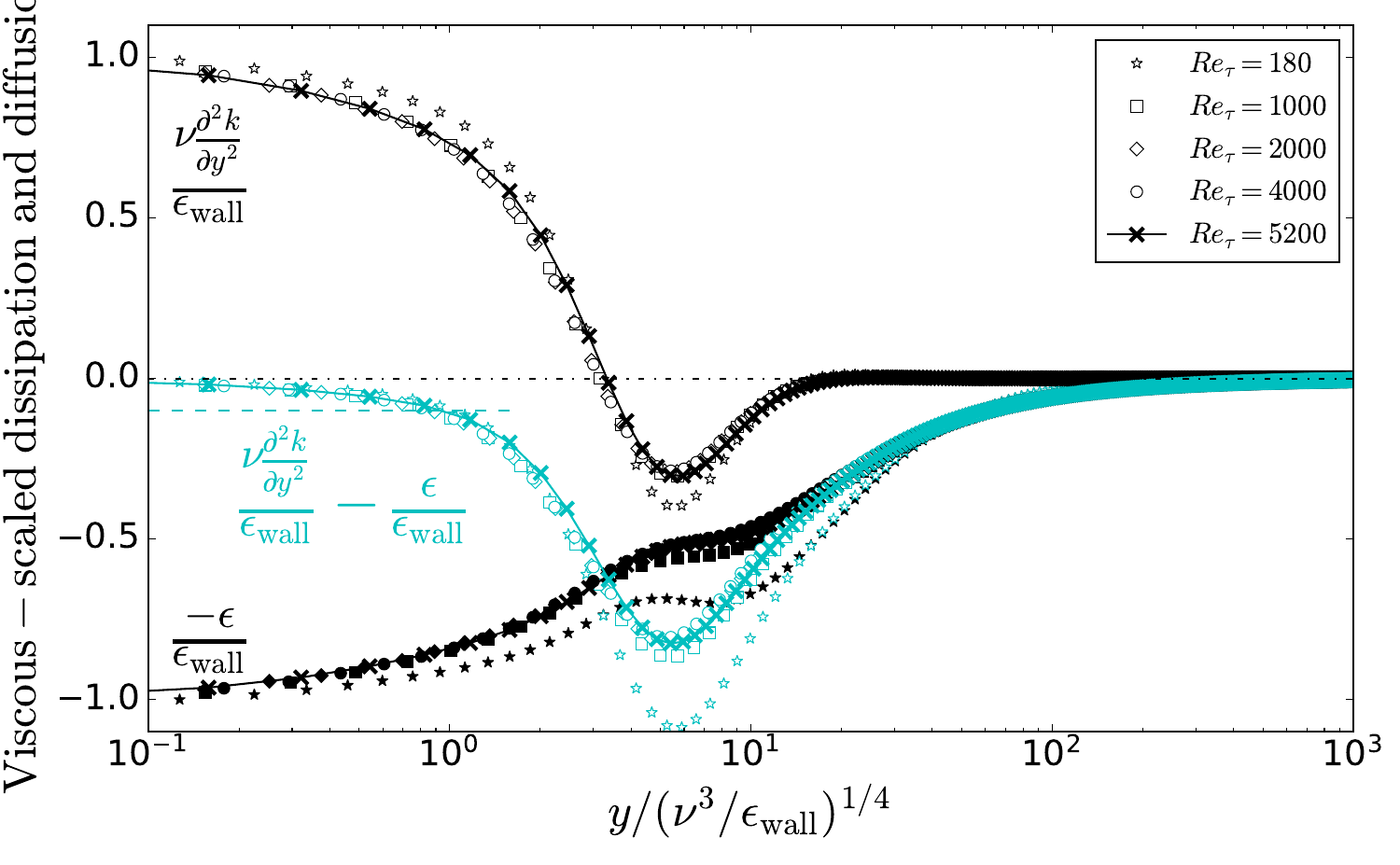}
		}
	}
	\vspace{-10pt}
	\caption{\it Scaling of the molecular diffusion and dissipation terms in the viscous sublayer, normalized by Kolmogorov scales defined using the dissipation rate at the wall. Data sources as in Figure \ref{fig:traditional-presentation-TKE-budget}. }
	\label{fig:viscous_scaling_TKE}
\end{figure}

Following a dimensional analysis of this region, Wei~\cite{wei2020scaling} proposed that the kinematic viscosity, $\nu$, and the wall dissipation rate, $\epsilon_\mathrm{wall}$, are the appropriate governing parameters for defining the characteristic length scale. This scale is equivalent to the Kolmogorov length evaluated using the wall dissipation rate, given by
\begin{equation}
	l_\mathrm{ref} = \left( \frac{\nu^3}{\epsilon_\mathrm{wall}} \right)^{1/4}.
\end{equation}
The corresponding scaled form of equation~(\ref{eq:scaling-viscous-sublayer}) is shown in figure~\ref{fig:viscous_scaling_TKE}, demonstrating a self-consistent collapse of data within the viscous sublayer.

The upper limit of the viscous sublayer can be estimated from the ratio of viscous diffusion to dissipation in figure~\ref{fig:ratio_over_dissipation}. The black dashed line marks the location where the ratio equals $-0.9$, indicating that approximately 90\% of the TKE balance is provided by viscous diffusion and dissipation, with the remaining 10\% due to other effects. Alternatively, the normalized terms in figure~\ref{fig:viscous_scaling_TKE} can be used to identify the upper limit of the viscous sublayer, where the cyan dashed line at a value of $-0.1$ denotes a wall-normal position of
\begin{equation}
	l_\mathrm{visc} \approx (\nu^3/\epsilon_\mathrm{wall})^{1/4}
\end{equation}
This result shows that the viscous sublayer associated with the TKE budget has a characteristic thickness on the order of the Kolmogorov length scale based on wall dissipation.

Since the dissipation rate is nearly impossible to be measured directly in experiments, it is useful to establish a relationship between the Kolmogorov length scale in the viscous sublayer and the conventional viscous length scale, $\nu/u_\tau$. The wall-normal distance normalized by the Kolmogorov scale can be expressed in terms of the inner-scaled distance as
	\begin{equation}
		\frac{y}{(\nu^3 / \epsilon_\mathrm{wall})^{1/4}} = (\epsilon^+_\mathrm{wall})^{1/4} \, y^+.
	\end{equation}
Direct numerical simulation data indicate that $\epsilon^+_\mathrm{wall}$ increases slowly with Reynolds number, following an approximately logarithmic trend. For example, at $Re_\tau = 1000$, $\epsilon^+_\mathrm{wall} \approx 0.252$, corresponding to $(\epsilon^+_\mathrm{wall})^{1/4} y^+ \approx 0.71 y^+$; whereas at $Re_\tau = 5200$, $\epsilon^+_\mathrm{wall} \approx 0.289$, yielding $(\epsilon^+_\mathrm{wall})^{1/4} y^+ \approx 0.81 y^+$. These results indicate that, for Reynolds numbers of practical relevance, the effective thickness of the viscous sublayer in the TKE budget equation can be approximated as $\Delta y^+_\mathrm{visc} \approx O(1)$.

\subsection{Scaling in the inner region}\label{section:inner-scaling}

\begin{figure}[h]
	\centering
	\vspace{3pt}
	\centerline{\hbox{ \hspace{-.0in} 
			\epsfxsize=6in
			\epsffile{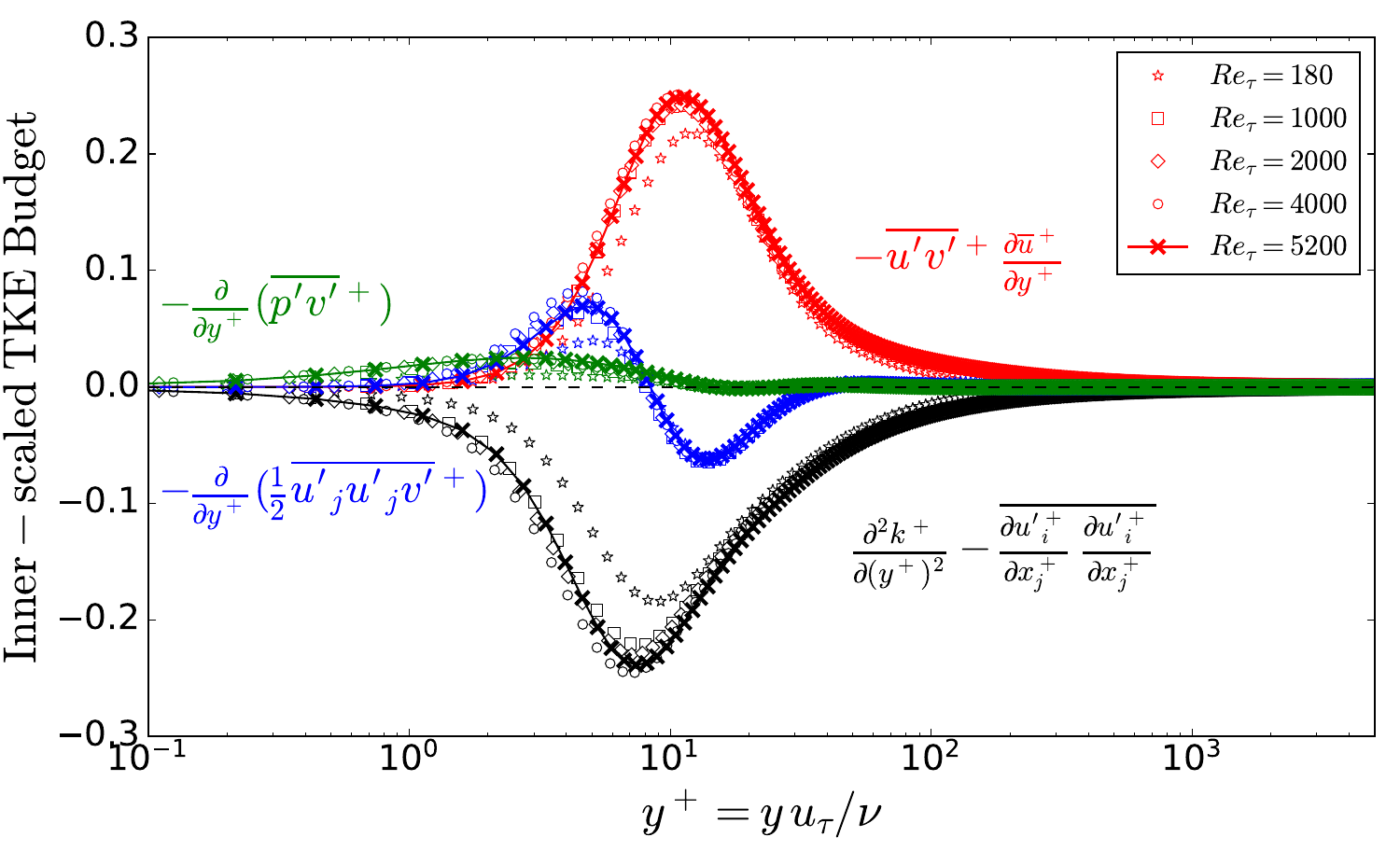}
		}
	}
	\vspace{-10pt}
	
	\caption{\it Inner scaling of the TKE budget terms using the reference scale $u^4_\tau/\nu$.  Data sources as in Figure \ref{fig:traditional-presentation-TKE-budget}. }
	\label{fig:inner_scaling_TKE}
\end{figure}

Figure~\ref{fig:traditional-presentation-TKE-budget} indicates that, outside the immediate viscous-sublayer, the TKE budget is primarily governed by a balance between the production and dissipation terms. A notable feature of the inner region is the distinct peak in the TKE production, which provides a natural reference scale for normalizing the TKE budget equation in this region. From the once-integrated mean momentum equation for fully developed turbulent channel flow, the Reynolds shear stress is given by
\begin{equation}
	- \overline{u' v'} =  u^2_\tau - u^2_\tau \frac{y}{\delta_c} - \nu \frac{\partial  \overline{u}}{\partial y}. 
\end{equation}
Accordingly, the TKE production term can be expressed as
\begin{equation}
		-\overline{u'v'} \, \frac{\partial  \overline{u}}{\partial  y} = \Big(u^2_\tau - u^2_\tau \frac{y}{\delta_c} - \nu \frac{\partial  \overline{u}}{\partial y} \Big)\frac{\partial  \overline{u}}{\partial  y}  
\end{equation}
Since the TKE production peak occurs close to the wall, where $y / \delta_c \ll O(1)$, the term involving $y / \delta_c$ may be neglected.   The peak production can then be estimated as
\begin{equation}
	\Big ( -\overline{u'v'} \, \frac{\partial  \overline{u}}{\partial  y} \Big)_\mathrm{max} \approx \Big (  \Big(  u^2_\tau  - \nu \frac{\partial  \overline{u}}{\partial y} \Big)\frac{\partial  \overline{u}}{\partial  y}\Big)_\mathrm{max}  \approx \frac{1}{4} \frac{u^4_\tau}{\nu},
\end{equation}
where the maximum occurs at $\partial  \overline{u} / \partial  y = 0.5 \, u_\tau^2 / \nu$.  Hence, $u_\tau^4 / \nu$ provides a natural reference scale, $A_\mathrm{ref}$, for the inner patch, corresponding to the traditional inner normalization of the TKE budget.   For clarity in data representation, it is convenient to group the two viscosity-related terms—viscous diffusion and dissipation—yielding the following dimensionless form of the TKE budget equation:
\begin{equation}\label{eq:inner-scaling}
	0 =  -\overline{u'v'}^+ \, \frac{\partial  \overline{u}^+}{\partial  y^+} + \Big ( \frac{\partial^2 k^+}{\partial  (y^+)^2} - \overline{ \frac{\partial u'^+_i}{\partial x^+_j} \frac{\partial u'^+_i}{\partial x^+_j}} \Big )   - \frac{\partial }{\partial y^+} \Big( \frac{1}{2} \overline{u'_j u'_j v'}^+  \Big) -   \frac{\partial }{\partial y^+} \Big( \overline{p' v'}^+ \Big)
\end{equation}
The inner-scaled TKE budget terms are presented in figure~\ref{fig:inner_scaling_TKE}, exhibiting good data collapse for $Re_\tau \gtrsim 1000$. The slight deviation observed at $Re_\tau = 180$ is attributed to low-Reynolds-number effects.

The upper limit of the inner layer can be defined as the location where the TKE production decreases to approximately 10\% of its peak value, which occurs around $y^+ \approx 100$. This estimate can be derived using the logarithmic law for the mean velocity profile, $u^+ \approx (1/\kappa)\ln(y^+) + B$, where $\kappa$ is the von Kármán constant and $B$ is an additive constant. Substituting this relation into the expression for the production term gives
\begin{equation}\label{eq:upper_edge_of_inner_layer}
	- \overline{u'v'}^+ \frac{\partial  \overline{u}^+}{\partial  y^+} = \Big(1 - \frac{\partial  \overline{u}^+}{\partial  y^+} - \frac{y^+}{\delta^+_c} \Big)\frac{\partial  \overline{u}^+}{\partial  y^+} \approx \frac{1}{\kappa} \frac{1}{y^+} - \Big(\frac{1}{\kappa} \frac{1}{y^+}\Big)^2 - \frac{1}{\kappa} \frac{1}{\delta^+_c},
\end{equation}
where the last term can be neglected for high-Reynolds-number flows. Taking $\kappa = 0.4$, equation~(\ref{eq:upper_edge_of_inner_layer}) yields a value of approximately $0.025$ at $y^+ \approx 100$, corresponding to about 10\% of the peak TKE production. Therefore, the extent of the inner patch can be reasonably estimated as $1 \lesssim y^+ \lesssim 100$.

\subsection{Scaling in the outer region}\label{section:outer-scaling}
In the outer region of turbulent channel flow, figure~\ref{fig:ratio_over_dissipation} shows that the viscous diffusion term becomes negligible. Consequently, the TKE budget equation can be approximated as
\begin{equation}
	0 \approx -\overline{u'v'} \, \frac{\partial  \overline{u}}{\partial  y}
	- \nu\, \overline{ \frac{\partial u'_i}{\partial x_j} \frac{\partial u'_i}{\partial x_j}}
	- \frac{\partial }{\partial y} \left( \frac{1}{2} \overline{u'_j u'_j v'} \right)
	- \frac{1}{\rho}\, \frac{\partial }{\partial y} \left(  \overline{p' v'} \right).
\end{equation}

Identifying appropriate reference scales for the dissipation and turbulent transport terms is not straightforward, as they involve spatial derivatives of velocity fluctuations and triple correlations. Therefore, we focus on the TKE production term to establish a suitable reference scale. Following the scaling-patch framework, once a consistent reference scale is determined for the production term, it can be applied analogously to the other terms in the TKE budget.

To estimate the reference scale for TKE production in the outer region, we invoke Prandtl’s mixing-length model~\cite{prandtl1925}, which states that	
\begin{equation}
	-\overline{u'v'} \sim \mathcal{l}^2 \Big(\frac{\partial  \overline{u}}{\partial  y}\Big)^2, \qquad \mathrm{or} \qquad \frac{\partial  \overline{u}}{\partial  y} \sim \frac{(-\overline{u'v'})^{1/2}}{\mathcal{l}}
\end{equation}
where $\mathcal{l}$ denotes the mixing length scale~\cite{prandtl1925, tennekes1972first}. From this, the TKE production can be expressed as
\begin{equation}
	-\overline{u'v'} \, \frac{\partial  \overline{u}}{\partial  y} \sim \frac{(-\overline{u'v'})^{3/2}}{\mathcal{l}}. 
\end{equation}
In the outer region of turbulent channel flow, the channel half-height, $\delta_c$, serves as a natural length scale. The Reynolds shear stress in this region decreases approximately linearly from its maximum value near the wall to zero at the channel centerline. Consequently, the maximum Reynolds shear stress provides a natural reference scale. It has been shown that $(-\overline{u'v'})_\mathrm{max} \approx u_\tau^2 \big(1 - O(1/\sqrt{Re_\tau})\big)$~\cite{fife2005stress}, indicating that $(-\overline{u'v'})_\mathrm{max} \approx u_\tau^2$ at sufficiently high Reynolds numbers of wall-bounded turbulent flows. Since the friction velocity $u_\tau$ is readily available in nearly all experimental and numerical studies of wall-bounded turbulence, it is convenient to define the reference scale for the TKE production term—and, by extension, for all other terms in the TKE budget—as $u_\tau^3 / \delta_c$. This reference scaling was also employed in the analysis of the TKE budget by Hoyas and Jiménez~\cite{hoyas2008reynolds}. Here, we derive this scaling directly from Prandtl’s mixing-length model and note that the same reasoning can be extended to other turbulent shear flows.

\begin{figure}[h]
	\centering
	\vspace{3pt}
	\centerline{\hbox{ \hspace{-.0in} 
			\epsfxsize=6in
			\epsffile{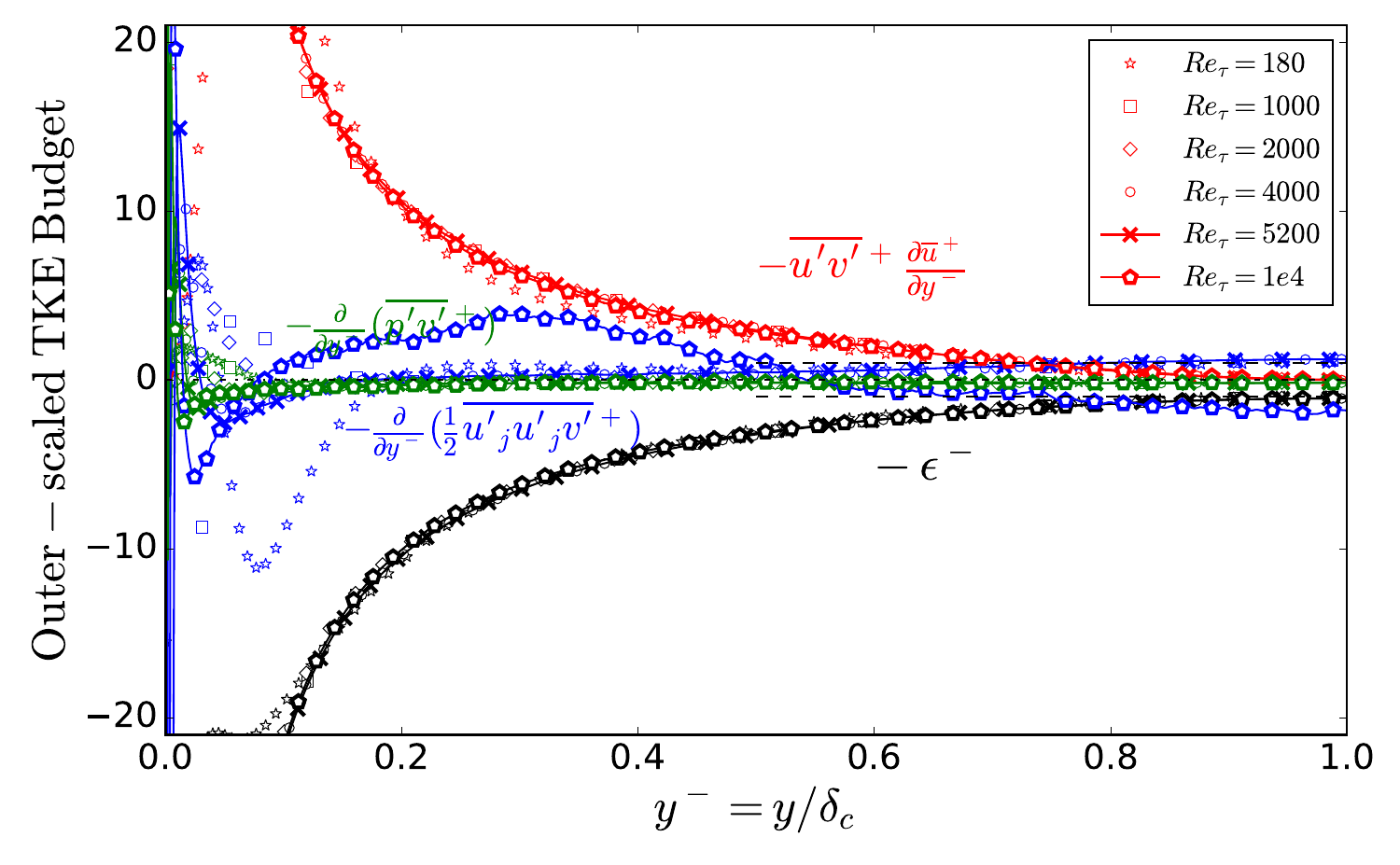}
		}
	}
	\vspace{-10pt}
	
	\caption{\it Outer scaling of TKE budget equation using $u^3_\tau/\delta_c$. Data sources as in Figure \ref{fig:traditional-presentation-TKE-budget}.}
	\label{fig:outer_scaling_TKE}
\end{figure}

Using $u_\tau^3 / \delta_c$ as the reference scale, the nondimensional TKE budget equation in the outer region becomes
\begin{equation}\label{eq:outer-scaling}
	0 \approx -\overline{u'v'}^+ \, \frac{\partial  \overline{u}^+}{\partial  y^-}
	\; - \; \epsilon^-
	\; - \; \frac{\partial }{\partial y^-} \left( \frac{1}{2} \overline{u'_j u'_j v'}^+ \right)
	\;-\;  \frac{\partial }{\partial y^-} \left(  \overline{p' v'}^+ \right),
\end{equation}
where $y^-=y/\delta_c$ and $\epsilon^- = \epsilon / (u_\tau^3 / \delta_c)$. Figure~\ref{fig:outer_scaling_TKE} shows the outer-scaled TKE budget terms, which exhibit consistent scaling behavior across different Reynolds numbers. For $y / \delta_c \gtrsim 0.2$, the scaled production and dissipation terms vary smoothly between approximately 0 and 10, demonstrating good collapse in the outer layer. As the channel centerline is approached, the TKE production term vanishes, while the dissipation is primarily balanced by the turbulent transport term. At the channel core, both the dissipation and turbulent transport terms attain a value of unity (highlighted by dashed line in the figure).

Figure \ref{fig:outer_scaling_TKE} confirms that the appropriate reference scale for the TKE production, dissipation, and other terms in the outer region is $u_\tau^3 / \delta_c$. For the dissipation term, if the velocity fluctuations $u'_i$ are normalized by $u_\tau$, then the corresponding length scale for spatial derivatives can be expressed as
	\begin{equation}
		(dx_j)_\mathrm{ref} = \left( \delta_c \, \frac{\nu}{u_\tau} \right)^{1/2}.
	\end{equation}
This expression represents a mixed length scale that incorporates both the outer length scale $\delta_c$ and the viscous length scale $\nu / u_\tau$. 
Notably, this mixed scale bears a strong resemblance to Taylor’s microscale, given by
	\begin{equation}
		\lambda = \left( \mathcal{l} \, \frac{\nu}{\mathcal{u}} \right)^{1/2},
	\end{equation}
where $\mathcal{l}$ denotes an integral length scale associated with the transfer of energy from the mean flow to turbulence, and $\mathcal{u}^2$ represents the kinetic energy of the large eddies~\cite{tennekes1972first}.

\subsection{Meso scaling of TKE budget}\label{section:meso-scaling}

Figure~\ref{fig:traditional-presentation-TKE-budget} shows that the inner-scaled dissipation, $\epsilon^+ = \epsilon / (u_\tau^4 / \nu)$, becomes much smaller than $O(1)$ for $y^+ \gtrsim 100$. Conversely, figure~\ref{fig:outer_scaling_TKE} demonstrates that the outer-scaled dissipation, $\epsilon^- = \epsilon / (u_\tau^3 / \delta_c)$, greatly exceeds $O(1)$ for $y/\delta_c \lesssim 0.1$. These observations indicate the existence of an intermediate, or meso, layer—possibly a hierarchy of meso-layers—within the TKE budget equation that bridges the inner and outer regions, approximately spanning the range $y^+ \gtrsim 100$ and $y/\delta_c \lesssim 0.1$.

It is worth noting that the inner and outer scalings defined in equations~(\ref{eq:inner-scaling}) and~(\ref{eq:outer-scaling}) are related through the Reynolds number as
\begin{equation}
	y^- = \frac{y^+}{Re_\tau}; \qquad \epsilon^- = \epsilon^+ Re_\tau. 
\end{equation}
This relationship suggests a scaling transition between the two asymptotic regions.

\begin{figure}[h]
	\centering
	\vspace{3pt}
	\centerline{\hbox{ \hspace{-.0in} 
			\epsfxsize=6in
			\epsffile{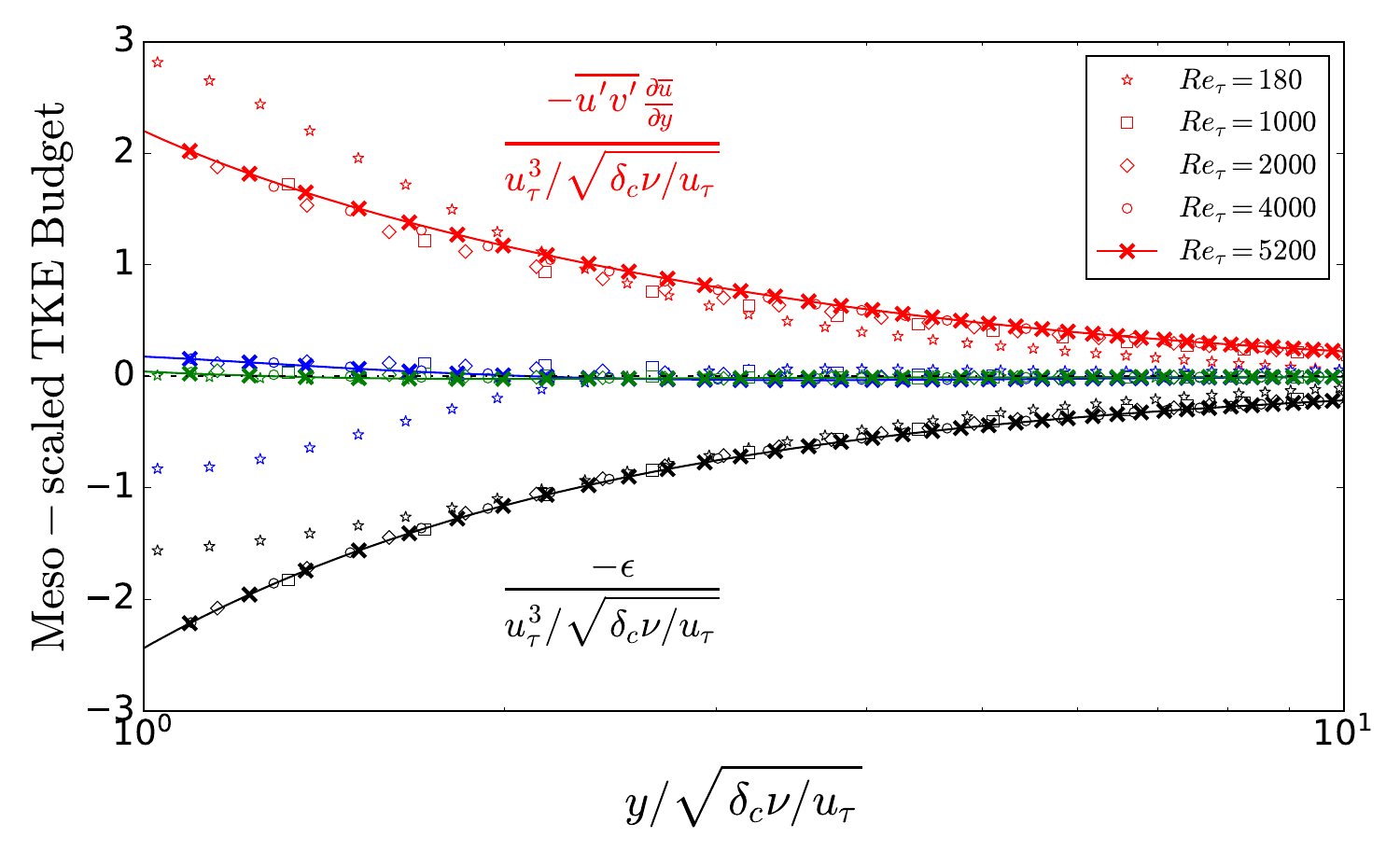}
		}
	}
	\vspace{-10pt}
	
	\caption{\it Meso scaling of TKE budget equation. Data sources as in Figure \ref{fig:traditional-presentation-TKE-budget}. }
	\label{fig:meso_scaling_TKE}
\end{figure}

In their scaling-patch analysis of the mean momentum equation, Wei et al.~\cite{wei2005properties} identified the natural emergence of a meso-layer characterized by a representative length scale $l_\mathrm{meso} = \sqrt{\delta_c \nu/u_\tau}$. While the concept of a meso-layer was first introduced by Afzal~\cite{afzal1982fully} and Long and Chen~\cite{long1981experimental}, the formulation by Wei et al. was developed within a distinct scaling-patch framework that offers a complementary perspective. Within this framework, the wall-normal coordinate in meso-scaling is defined as
\begin{equation}
	y^* = \frac{y}{l_\mathrm{meso}} = \frac{y}{\sqrt{\delta_c \nu/u_\tau}} = \frac{y^+}{\sqrt{Re_\tau}} = y^- \sqrt{Re_\tau}.
\end{equation}
Building on this concept, we extend the meso-scaling framework to the TKE budget equation by introducing a reference scale proportional to $Re^{-1/2}_\tau (u_\tau^4 / \nu)$. Accordingly, the dimensionless form of the TKE budget can be obtained either by multiplying the inner-scaled equation~(\ref{eq:inner-scaling}) by $(Re_\tau)^{1/2}$ or, equivalently, by dividing the outer-scaled equation~(\ref{eq:outer-scaling}) by \((Re_\tau)^{1/2}\). The meso-scaled TKE production and dissipation are then
\begin{equation}
	\Big(-\overline{u'v'} \, \frac{\partial  \overline{u}}{\partial  y}\Big)^* = \frac{-\overline{u'v'} \, \frac{\partial  \overline{u}}{\partial  y}}{u^3_\tau/\sqrt{\delta_c \nu/u_\tau}} = \frac{-\overline{u'v'} \, \frac{\partial  \overline{u}}{\partial  y}}{u^4_\tau/(\nu \sqrt{Re_\tau})};  \qquad \epsilon^* = \frac{\epsilon}{u^3_\tau/\sqrt{\delta_c \nu/u_\tau}}.
\end{equation}
The resulting meso-scaled TKE budget, shown in figure~\ref{fig:meso_scaling_TKE}, captures the intermediate scaling behavior between the inner and outer regions.

If the proposed meso-scaling appropriately represents the dynamics in this region, then the scaled TKE budget terms should be of \(O(1)\) within the meso-layer, defined by \(y^* = O(1)\). Figure~\ref{fig:meso_scaling_TKE} indicates that this corresponds to \(1 \lesssim y^* \lesssim 10\). For \(Re_\tau = 5200\), this range translates to \(72 < y^+ < 721\) or \(0.01 < y^- < 0.14\); for \(Re_\tau = 180\), it corresponds to \(13 < y^+ < 134\) or \(0.07 < y^- < 0.7\).

\section{Application to the TKE budget equation in turbulent boundary layer flow over a flat plate }\label{section:ZPGTBL}
For a turbulent boundary layer over a flat plate, or zero-pressure-gradient turbulent boundary layer (ZPG TBL),  the TKE budget equation is \cite{tennekes1972first}
\begin{align}\label{eq:TKE_budget_ZPG_TBL}
0 =&  \underbrace{\Big \{ -\overline{u'v'}\frac{\partial \overline{u}}{\partial y} \; - \overline{u'v'}\frac{\partial \overline{v}}{\partial x} \;  - \overline{u'u'}\frac{\partial \overline{u}}{\partial x}  \;  - \overline{v'v'}\frac{\partial \overline{v}}{\partial y} \Big\}}_{\mathrm{prod.}} \; + \; 	\underbrace{\Big\{- \nu\, \overline{ \frac{\partial u'_i}{\partial x_j} \frac{\partial u'_i}{\partial x_j}} \Big\} }_{\mathrm{diss. \, or\, }-\epsilon} \; + \; 	\underbrace{\Big\{ \nu \frac{\partial^2 k}{\partial x^2} + \nu \frac{\partial^2 k}{\partial y^2}\Big\}}_{\mathrm{visc\_diff.}} \nonumber \\[10pt]
& \; + \; 	\underbrace{\Big\{ - \frac{\partial}{\partial x} \Big( \frac{1}{2} \overline{u'_j u'_j u'}  \Big) - \frac{\partial}{\partial y} \Big( \frac{1}{2} \overline{u'_j u'_j v'}  \Big) \Big\}}_{\mathrm{turb\_tran.}} \; + \; 	\underbrace{\Big\{ - \frac{1}{\rho}\,\frac{\partial}{\partial x} \Big(  \overline{p' u'} \Big) - \frac{1}{\rho}\,\frac{\partial}{\partial y} \Big(  \overline{p' v'} \Big) \Big\}}_{\mathrm{pres\_tran.}}  \nonumber \\[10pt]
&\; + \;\underbrace{-\overline{u} \frac{\partial k}{\partial x} - \overline{v} \frac{\partial k}{\partial y}}_{\mathrm{conv.}}. 
\end{align}

Compared with the channel flow formulation in equation~(\ref{eq:TKE_budget}), the TKE budget for a ZPG TBL includes an additional mean-convection term (the last term on the right-hand side of equation (\ref{eq:TKE_budget_ZPG_TBL})). Moreover, because the ZPG TBL evolves in the streamwise ($x$) direction, the production, turbulent transport, viscous diffusion, and pressure–velocity correlation terms each acquire additional components involving streamwise derivatives. For instance, by denoting the TKE production term in a ZPG TBL as $A$ in the generic form of equation~(\ref{eq:generic_equation_A}), one obtains
	\begin{equation}
		A = -\overline{u'v'}\frac{\partial \overline{u}}{\partial y} \; - \overline{u'v'}\frac{\partial \overline{v}}{\partial x} \;  - \overline{u'u'}\frac{\partial \overline{u}}{\partial x}  \;  - \overline{v'v'}\frac{\partial \overline{v}}{\partial y}.
	\end{equation}
If a reference scale $A_\mathrm{ref}$ is chosen, the normalized production term becomes
	\begin{equation}
		A^* = \frac{A}{A_\mathrm{ref}} =  \big ( A^*_\mathrm{I} + A^*_\mathrm{II} + A^*_\mathrm{III} + A^*_\mathrm{IV} \big).
	\end{equation}
For instance, using the outer reference for TKE production introduced in Section~\ref{section:inner-scaling}, $A_\mathrm{ref} = u_\tau^3/\delta_e$, the leading contribution $A^*_\mathrm{I}$ corresponds to that in channel flow. In a ZPG TBL, the boundary layer is thin, variations in the $x$-direction are much smaller than those in the $y$-direction, and the mean wall-normal velocity is much smaller than the streamwise velocity. Consequently, the additional production components $A^*_\mathrm{II}$, $A^*_\mathrm{III}$, and $A^*_\mathrm{IV}$ are all $\ll O(1)$. Thus, the scaling of TKE production in ZPG TBLs is equivalent to that in turbulent channel flow. Similarly, the scalings of the turbulent-transport, viscous-diffusion, and pressure–velocity correlation terms are also the same as those in turbulent channel flow, despite the presence of additional $x$-derivative terms.

In the following analysis, the scaling relationships established for channel flow are applied to DNS data of ZPG TBLs at momentum-thickness Reynolds numbers $Re_\theta = 670$, $1000$, $2000$, $3030$, and $4060$, corresponding to friction Reynolds numbers $Re_\tau = 250$, $360$, $670$, $970$, and $1270$, respectively. Due to computational constraints, these simulations are limited to moderate Reynolds numbers.

\subsection{Scaling in viscous sublayer in ZPG TBL}

\begin{figure}[h]
	\centering
	\vspace{3pt}
	\centerline{\hbox{ \hspace{-.0in} 
			\epsfxsize=5in
			\epsffile{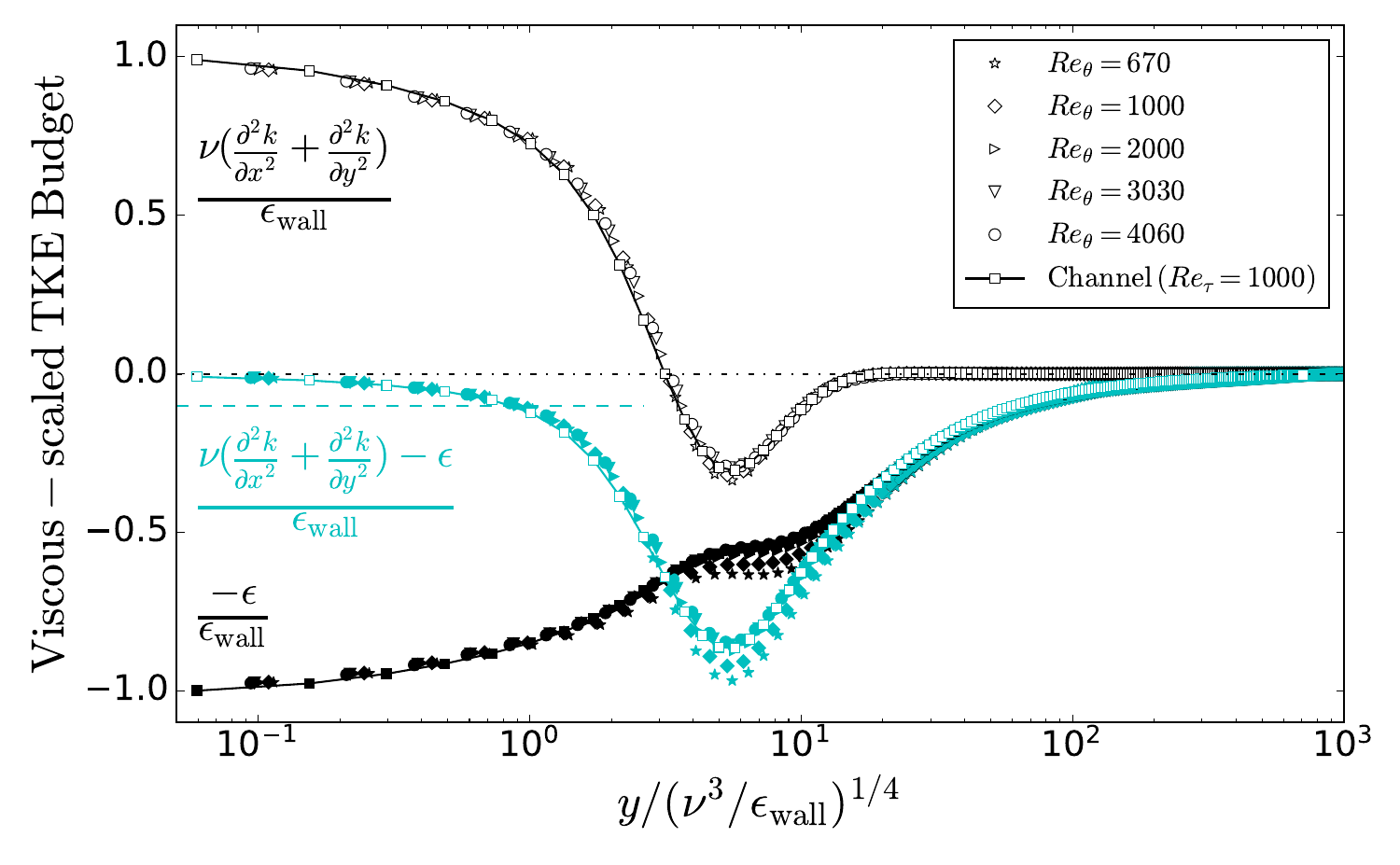}
		}
	}
	\vspace{-10pt}
	
	\caption{\it Scaling of viscous diffusion and dissipation in the viscous sublayer of ZPG TBLs. The ZPG TBL data are obtained from the DNS of Schlatter and \"{O}rl\"{u} \cite{schlatter2010assessment}, while the turbulent channel flow data at $Re_\tau = 1000$ are from the DNS of Bernardini et al. \cite{bernardini2014velocity}.  }
	\label{fig:ZPGTBL_viscous}
\end{figure}

Figure~\ref{fig:ZPGTBL_viscous} shows the viscous diffusion and dissipation terms in the viscous sublayer, scaled according to the formulation developed in Section~\ref{section:viscous-scaling}. For reference, DNS data of turbulent channel flow at $Re_\tau = 1000$ are also included. The figure demonstrates that the channel-flow-derived scaling successfully collapses the ZPG TBL data, particularly in the near-wall region, where viscous diffusion and dissipation exhibit nearly identical behavior in both flows. This close agreement supports the universality of the viscous scaling across canonical wall-bounded turbulent flows.

\subsection{Inner scaling of TKE budget in ZPG TBL}

\begin{figure}[h]
	\centering
	\vspace{3pt}
	\centerline{\hbox{ \hspace{-.0in} 
			\epsfxsize=5in
			\epsffile{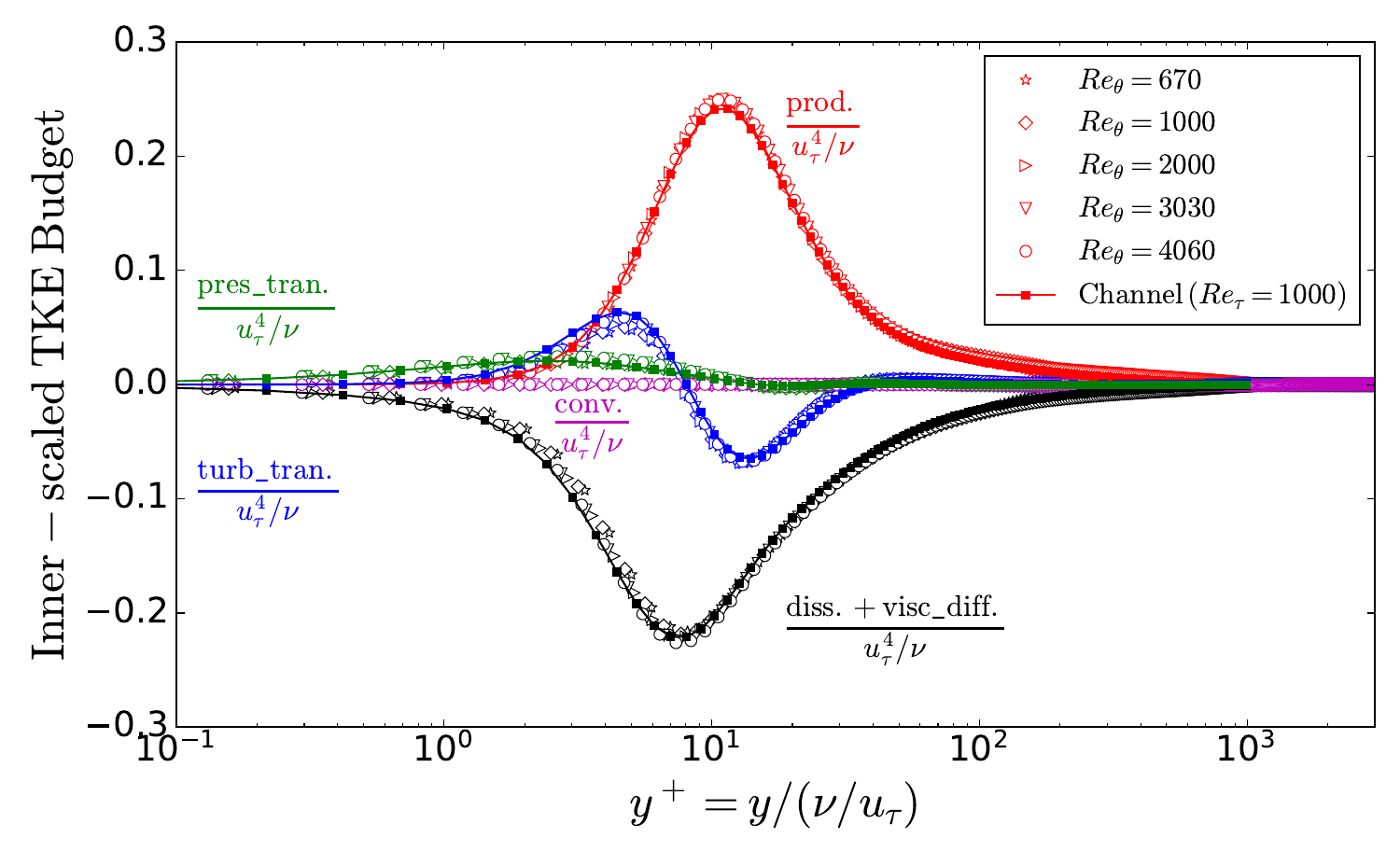}
		}
	}
	\vspace{-10pt}
	
	\caption{\it Scaling of the TKE budgets of ZPG TBLs in the inner layer. The magenta $\mathrm{conv}^+$ denotes the inner scaled mean convection term. Data sources as in Figure~\ref{fig:ZPGTBL_viscous}.  }
	\label{fig:ZPGTBL_inner}
\end{figure}

Figure~\ref{fig:ZPGTBL_inner} presents the TKE budget of ZPG TBLs scaled using the formulation developed in Section~\ref{section:inner-scaling}. In this inner region, the conventionally scaled TKE budget exhibits trends that closely match those observed in turbulent channel flow, confirming the consistency of the inner-layer scaling between these two canonical wall-bounded flows.

\subsection{Outer scaling of TKE budget in ZPG TBL}

\begin{figure}[h]
	\centering
	\vspace{3pt}
	\centerline{\hbox{ \hspace{-.0in} 
			\epsfxsize=5in
			\epsffile{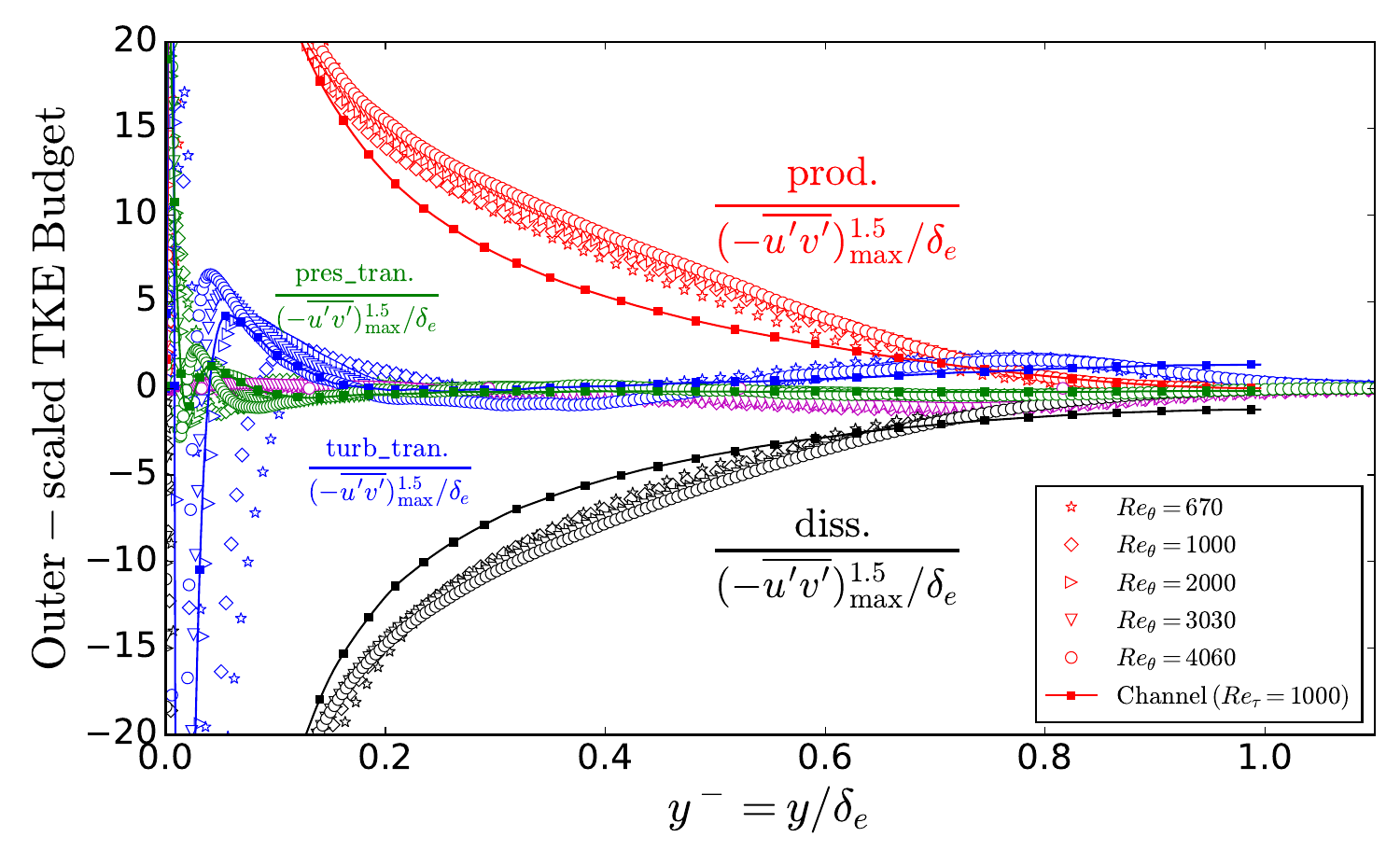}
		}
	}
	\vspace{-10pt}
	
	\caption{\it Outer-scaling of the TKE budgets in ZPG TBLs. Data sources as in Figure~\ref{fig:ZPGTBL_viscous}.   }
	\label{fig:ZPGTBL_outer_use_Ruvmax}
\end{figure}

Figure~\ref{fig:ZPGTBL_outer_use_Ruvmax} presents the outer-scaled TKE budget of ZPG TBLs. Given the relatively moderate Reynolds-number range, the reference scale is chosen as $(-\overline{u'v'})_\mathrm{max}^{1.5}/\delta_c$ instead of $u_\tau^3/\delta_c$. Since the deviation of $(-\overline{u'v'})_\mathrm{max}$ from $u_\tau^2$ is not negligible at these Reynolds numbers, using $(-\overline{u'v'})_\mathrm{max}^{1.5}/\delta_c$ yields a better collapse of the TKE budget terms in the outer region. Unlike the viscous sublayer and inner layer, the outer region exhibits clear distinctions between ZPG TBLs and turbulent channel flow, reflecting the fundamentally different flow environments near the boundary-layer edge and the channel centerline.

\begin{figure}[h]
	\centering
	\vspace{3pt}
	\centerline{\hbox{ \hspace{-.0in} 
			\epsfxsize=6in
			\epsffile{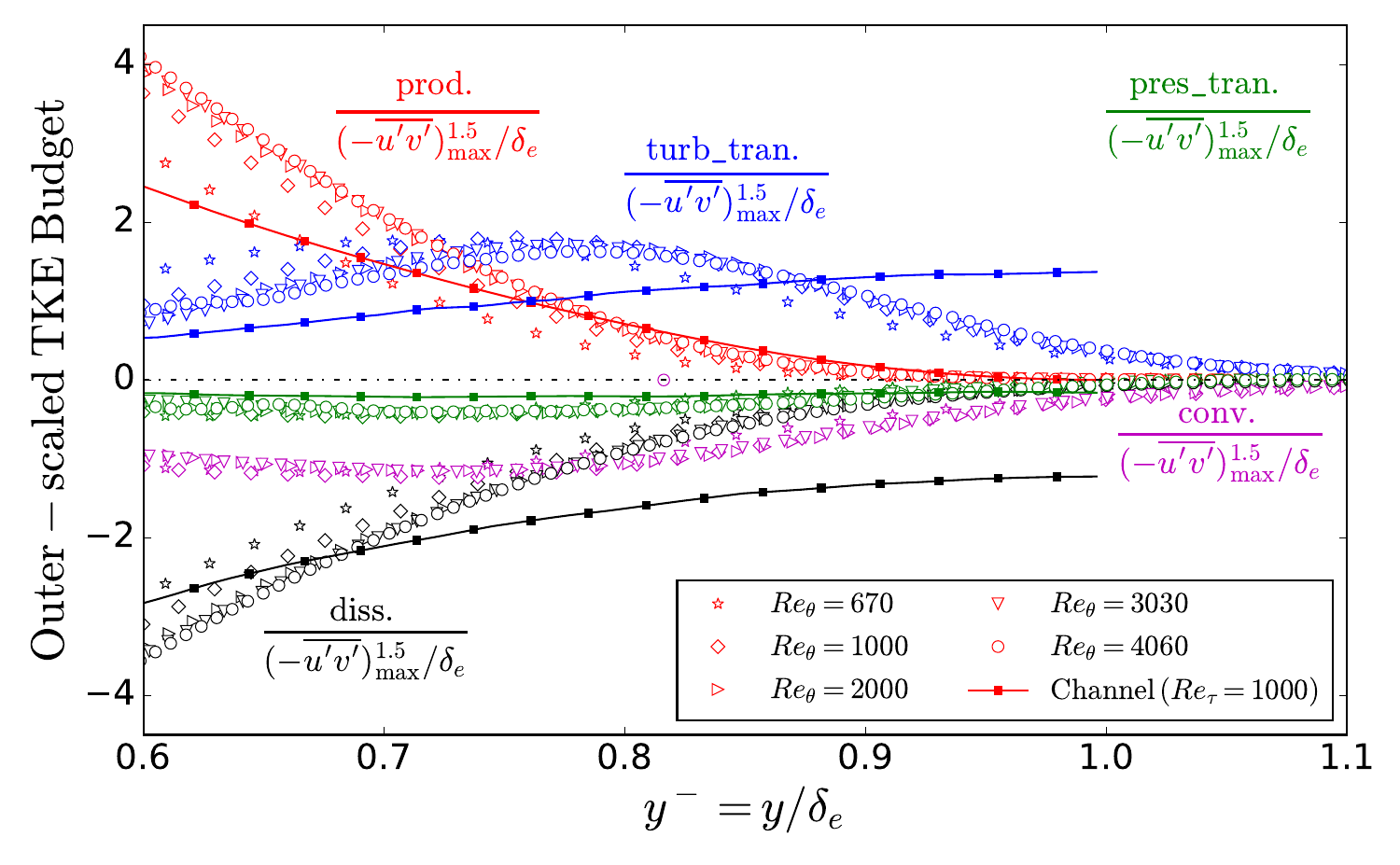}
		}
	}
	\vspace{-10pt}
	
	\caption{\it Outer-layer scaling of TKE budgets in ZPG TBLs, with a focus on the outer region. Data sources as in Figure~\ref{fig:ZPGTBL_viscous}.   }
	\label{fig:ZPGTBL_outer_use_Ruvmax_zoomin}
\end{figure}

To further examine these outer-region differences, Figure~\ref{fig:ZPGTBL_outer_use_Ruvmax_zoomin} focuses on the range $y/\delta_c \geq 0.6$, where the distinctions become more evident. In ZPG TBLs, both TKE production and dissipation diminish toward the boundary-layer edge, where the TKE budget is mainly balanced by turbulent transport and mean convection. In contrast, in turbulent channel flow, dissipation remains finite at the centerline and is balanced predominantly by the turbulent transport term.

\subsection{Meso-scaling of TKE budget in ZPG TBL}
\begin{figure}[h]
	\centering
	\vspace{3pt}
	\centerline{\hbox{ \hspace{-.0in} 
			\epsfxsize=6in
			\epsffile{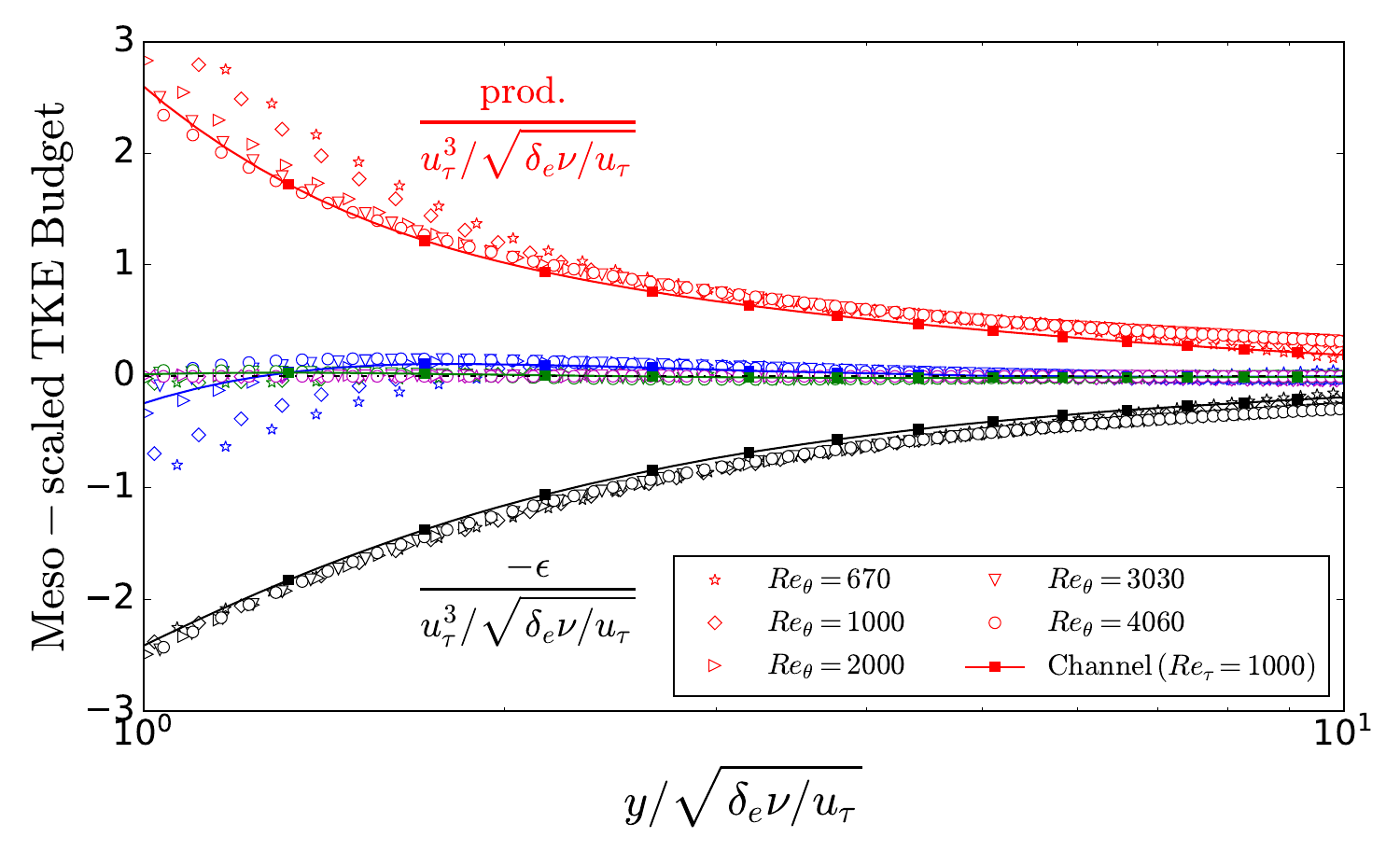}
		}
	}
	\vspace{-10pt}
	
	\caption{\it Meso-scaling of TKE budgets in ZPG TBLs. Data sources as in Figure~\ref{fig:ZPGTBL_viscous}.   }
	\label{fig:ZPGTBL_meso_scaling}
\end{figure}

Figure~\ref{fig:ZPGTBL_meso_scaling} presents the meso-scaled TKE budget for the ZPG TBL. The data exhibit a consistent collapse across the meso-layer, demonstrating the robustness of the proposed scaling. Both production and dissipation remain of order unity and vary smoothly throughout this region, confirming that the meso-scaling appropriately captures the dominant balance. In contrast, viscous, turbulent, and pressure transport terms contribute negligibly to the energy balance within the meso-layer.

\clearpage
\section{Summary}

The present study extends the scaling-patch framework  to the turbulent kinetic energy (TKE) budget in wall-bounded turbulent flows. The analysis identifies distinct scaling patches, each governed by a dominant balance among the TKE budget terms and characterized by its own natural scaling parameters. Within each patch, the appropriately normalized variables remain order unity and vary gradually, enabling a unified and physically consistent description of the TKE budget across the flow.

The analysis reformulates the TKE budget by introducing a lumped reference scale for the dominant terms in each sublayer. This approach enables systematic identification of dominant balances without prescribing individual scales for each constituent quantity. A scaling patch is defined as a region where at least two terms in the TKE equation are of comparable magnitude, $O(1)$, while the remaining terms are asymptotically small. The relative strengths of production, dissipation, viscous diffusion, turbulent transport, and pressure diffusion delineate the near-wall, meso-, and outer-layer regions.

In the viscous sublayer, the TKE budget is dictated by a local balance between viscous diffusion and dissipation. The corresponding reference scale arises naturally from the kinematic viscosity and the wall dissipation rate, and the sublayer thickness is of the order of the Kolmogorov length scale.

Within the inner region, the peak of TKE production provides a natural reference, leading to the conventional inner scaling $u_\tau^4/\nu$. When the viscous diffusion and dissipation terms are grouped under this scaling, a markedly improved collapse across different Reynolds numbers is obtained. The upper limit of the inner layer is found to occur near $y^+ \approx 100$.

In the outer region, where viscous effects are negligible, Prandtl’s mixing-length model yields the appropriate reference scaling $u_\tau^3/\delta_c$. To bridge the inner and outer regimes, a meso-scaling of $u_\tau^3/\sqrt{\delta_c \nu / u_\tau}$ was introduced, with the wall-normal distance normalized by the meso-length scale $l_\mathrm{meso} = \sqrt{\delta_c \nu / u_\tau}$. The resulting dimensionless TKE budget reveals an intermediate regime, spanning approximately $y^+ \sim 10^2$ to $y/\delta_c \sim 10^{-1}$.

This unified, multi-layer framework delineates the smooth transition among viscous, inner, meso, and outer dynamics for the TKE budget. It provides not only a cohesive interpretation of the underlying physical mechanisms but also a systematic foundation for extending the scaling-patch approach to other multiscale turbulent transport equations.

\vspace{0.2in}
\section*{ACKNOWLEDGMENT}
The authors are grateful to Dr. M. Lee, Dr. R. D. Moser, Dr. P. Schlatter, and Dr. \"{O}rl\"{u} for generously providing access to their DNS data.

\bibliography{./ref_tke_scaling}

\end{document}